\documentclass{article}
\usepackage{graphicx}
\usepackage{amsmath}
\usepackage{amsfonts}
\usepackage{amssymb}
\input epsf
\advance \topmargin by -\headheight
\advance \topmargin by -\headsep
\evensidemargin \oddsidemargin
\begin{document}

\title{The diagonalization of quantum field Hamiltonians}
\author{Dean Lee$^{a}\thanks{dlee@physics.umass.edu}$, Nathan Salwen$^{b}%
\thanks{salwen@physics.harvard.edu}$, and Daniel Lee$^{c}%
\thanks{ddlee@bell-labs.com}$\\$^{a}$Dept. of Physics, Univ. of Massachusetts, Amherst, MA 01003\\$^{b}$Dept. of Physics, Harvard Univ., Cambridge, MA 02138\\$^{c}$Physical Sciences Division, Bell Laboratories, NJ 07974}
\maketitle
\begin{abstract}
We introduce a new diagonalization method called quasi-sparse eigenvector
diagonalization which finds the most important basis vectors of the low energy
eigenstates of a quantum Hamiltonian. \ It can operate using any basis, either
orthogonal or non-orthogonal, and any sparse Hamiltonian, either Hermitian,
non-Hermitian, finite-dimensional, or infinite-dimensional. \ The method is
part of a new computational approach which combines both diagonalization and
Monte Carlo techniques. \ 
\end{abstract}

\section{Introduction}

Most computational work in non-perturbative quantum field theory and many body
phenomena rely on one of two general techniques, Monte Carlo or
diagonalization. \ These methods are nearly opposite in their strengths and
weaknesses. \ Monte Carlo requires relatively little storage, can be performed
using parallel processors, and in some cases the computational effort scales
reasonably with system size. \ But it has great difficulty for systems with
sign or phase oscillations and provides only indirect information on
wavefunctions and excited states. \ In contrast diagonalization methods do not
suffer from fermion sign problems, can handle complex-valued actions, and can
extract details of the spectrum and eigenstate wavefunctions. \ However the
main problem with diagonalization is that the required memory and CPU time
scales exponentially with the size of the system.

In view of the complementary nature of the two methods, we consider the
combination of both diagonalization and Monte Carlo within a computational
scheme. \ We propose a new approach which takes advantage of the strengths of
the two computational methods in their respective domains. \ The first half of
the method involves finding and diagonalizing the Hamiltonian restricted to an
optimal subspace. \ This subspace is designed to include the most important
basis vectors of the lowest energy eigenstates. \ Once the most important
basis vectors are found and their interactions treated exactly, Monte Carlo is
used to sample the contribution of the remaining basis vectors. \ By this
two-step procedure much of the sign problem is negated by treating the
interactions of the most important basis states exactly, while storage and CPU
problems are resolved by stochastically sampling the collective effect of the
remaining states.

In our approach diagonalization is used as the starting point of the Monte
Carlo calculation. \ Therefore the two methods should not only be efficient
but work well together. \ On the diagonalization side there are several
existing methods using Tamm-Dancoff truncation \cite{perry}, similarity
transformations \cite{wilson}, density matrix renormalization group
\cite{white}, or variational algorithms such as stochastic diagonalization
\cite{husslein}. \ However we find that each of these methods is either not
sufficiently general, not able to search an infinite or large dimensional
Hilbert space, not efficient at finding important basis vectors, or not
compatible with the subsequent Monte Carlo part of the calculation. \ The
Monte Carlo part of our diagonalization/Monte Carlo scheme is discussed
separately in a companion paper \cite{sec}. \ In this paper we consider the
diagonalization part of the scheme. \ We introduce a new diagonalization
method called quasi-sparse eigenvector (QSE) diagonalization. \ It is a
general algorithm which can operate using any basis, either orthogonal or
non-orthogonal, and any sparse Hamiltonian, either real, complex, Hermitian,
non-Hermitian, finite-dimensional, or infinite-dimensional. \ It is able to
find the most important basis states of several low energy eigenvectors
simultaneously, including those with identical quantum numbers, from a random
start with no prior knowledge about the form of the eigenvectors.

Our discussion is organized as follows. \ We first define the notion of
quasi-sparsity in eigenvectors and introduce the quasi-sparse eigenvector
method. \ We discuss when the low energy eigenvectors are likely to be
quasi-sparse and make an analogy with Anderson localization. \ We then
consider three examples which test the performance of the algorithm. \ In the
first example we find the lowest energy eigenstates for a random sparse real
symmetric matrix. \ In the second example we find the lowest eigenstates
sorted according to the real part of the eigenvalue for a random sparse
complex non-Hermitian matrix. \ In the last example we consider the case of an
infinite-dimensional Hamiltonian defined by $1+1$ dimensional $\phi^{4}$
theory in a periodic box. \ We conclude with a summary and some comments on
the role of quasi-sparse eigenvector diagonalization within the context of the
new diagonalization/Monte Carlo approach.

\section{Quasi-sparse eigenvector method}

Let $\left|  e_{i}\right\rangle $ denote a complete set of basis vectors.
\ For a given energy eigenstate
\begin{equation}
|v\rangle=\sum_{i}c_{i}\left|  e_{i}\right\rangle ,
\end{equation}
we define the important basis states of $|v\rangle$ to be those $\left|
e_{i}\right\rangle $ such that for fixed normalizations of $|v\rangle$ and the
basis states, $\left|  c_{i}\right|  $ exceeds a prescribed threshold value.
\ If $|v\rangle$ can be well-approximated by the contribution from only its
important basis states we refer to the eigenvector $|v\rangle$ as
\textit{quasi-sparse} with respect to $\left|  e_{i}\right\rangle $.

Standard sparse matrix algorithms such as the Lanczos or Arnoldi methods allow
one to find the extreme eigenvalues and eigenvectors of a sparse matrix
efficiently, without having to store or manipulate large non-sparse matrices.
\ However in quantum field theory or many body theory one considers very large
or infinite dimensional spaces where even storing the components of a general
vector is impossible. \ For these more difficult problems the strategy is to
approximate the low energy eigenvectors of the large space by diagonalizing
smaller subspaces. \ If one has sufficient intuition about the low energy
eigenstates it may be possible to find a useful truncation of the full vector
space to an appropriate smaller subspace. \ In most cases, however, not enough
is known \textit{a priori }about the low energy eigenvectors. \ The dilemma is
that to find the low energy eigenstates one must truncate the vector space,
but in order to truncate the space something must be known about the low
energy states.

Our solution to this puzzle is to find the low energy eigenstates and the
appropriate subspace truncation at the same time by a recursive process. \ We
call the method quasi-sparse eigenvector (QSE) diagonalization, and we
describe the steps of the algorithm as follows. \ The starting point is any
complete basis for which the Hamiltonian matrix $H_{ij}$ is sparse. \ The
basis vectors may be non-orthogonal and/or the Hamiltonian matrix may be
non-Hermitian. \ The following steps are now iterated:

\begin{enumerate}
\item  Select a subset of basis vectors $\left\{  e_{i_{1}},\cdots,e_{i_{n}%
}\right\}  $ and call the corresponding subspace $S$.

\item  Diagonalize $H$ restricted to $S$ and find one eigenvector $v$.

\item  Sort the basis components of $v$ according to their magnitude and
remove the least important basis vectors.

\item  Replace the discarded basis vectors by new basis vectors. \ These are
selected at random according to some weighting function from a pool of
candidate basis vectors which are connected to the old basis vectors through
non-vanishing matrix elements of $H$.

\item  Redefine $S$ as the subspace spanned by the updated set of basis
vectors and repeat steps 2 through 5.
\end{enumerate}

If the subset of basis vectors is sufficiently large, the exact low energy
eigenvectors will be stable fixed points of the QSE update process. \ We can
show this as follows. \ Let $\left|  i\right\rangle $ be the eigenvectors of
the submatrix of $H$ restricted to the subspace $S$, where $S$ is the span of
the subset of basis vectors after step 3 of the QSE algorithm. \ Let $\left|
A_{j}\right\rangle $ be the remaining basis vectors in the full space not
contained in $S$. \ We can represent $H$ as
\begin{equation}
\left[
\begin{array}
[c]{cccccc}%
\lambda_{1} & 0 & \cdots & \left\langle 1\right|  H\left|  A_{1}\right\rangle
& \left\langle 1\right|  H\left|  A_{2}\right\rangle  & \cdots\\
0 & \lambda_{2} & \cdots & \left\langle 2\right|  H\left|  A_{1}\right\rangle
& \left\langle 2\right|  H\left|  A_{2}\right\rangle  & \cdots\\
\vdots & \vdots & \ddots & \vdots & \vdots & \cdots\\
\left\langle A_{1}\right|  H\left|  1\right\rangle  & \left\langle
A_{1}\right|  H\left|  2\right\rangle  & \cdots &  E\cdot\lambda_{A_{1}} &
\left\langle A_{1}\right|  H\left|  A_{2}\right\rangle  & \cdots\\
\left\langle A_{2}\right|  H\left|  1\right\rangle  & \left\langle
A_{2}\right|  H\left|  2\right\rangle  & \cdots & \left\langle A_{2}\right|
H\left|  A_{1}\right\rangle  & E\cdot\lambda_{A_{2}} & \cdots\\
\vdots & \vdots & \vdots & \vdots & \vdots & \ddots
\end{array}
\right]  . \label{matrix}%
\end{equation}
We have used Dirac's bra-ket notation to represent the terms of the
Hamiltonian matrix. \ In cases where the basis is non-orthogonal and/or the
Hamiltonian is non-Hermitian, the meaning of this notation may not be clear.
\ When writing $\left\langle A_{1}\right|  H\left|  1\right\rangle $, for
example, we mean the result of the dual vector to $\left|  A_{1}\right\rangle
$ acting upon the vector $H\left|  1\right\rangle $. \ In (\ref{matrix}) we
have written the diagonal terms for the basis vectors $\left|  A_{j}%
\right\rangle $ with an explicit factor $E$. \ We let $\left|  1\right\rangle
$ be the approximate eigenvector of interest and have shifted the diagonal
entries so that $\lambda_{1}=0.$ \ Our starting hypothesis is that $\left|
1\right\rangle $ is close to some exact eigenvector of $H$ which we denote as
$\left|  1_{\text{full}}\right\rangle $. \ More precisely we assume that the
components of $\left|  1_{\text{full}}\right\rangle $ outside $S$ are small
enough so that we can expand in inverse powers of the introduced parameter $E.$

We now expand the eigenvector as
\begin{equation}
\left|  1_{\text{full}}\right\rangle =\left[
\begin{array}
[c]{c}%
1\\
c_{2}^{\prime}E^{-1}+\cdots\\
\vdots\\
c_{A_{1}}^{\prime}E^{-1}+\cdots\\
c_{A_{2}}^{\prime}E^{-1}+\cdots\\
\vdots
\end{array}
\right]  \label{eigvec}%
\end{equation}
and the corresponding eigenvalue as
\begin{equation}
\lambda_{\text{full}}=\lambda_{1}^{\prime}E^{-1}+\cdots.
\end{equation}
In (\ref{eigvec}) we have chosen the normalization of $\left|  1_{\text{full}%
}\right\rangle $ such that $\left\langle 1\right.  \left|  1_{\text{full}%
}\right\rangle =1$. \ From the eigenvalue equation
\begin{equation}
H\left|  1_{\text{full}}\right\rangle =\lambda_{\text{full}}\left|
1_{\text{full}}\right\rangle
\end{equation}
we find at lowest order
\begin{equation}
c_{A_{j}}^{\prime}=-\tfrac{\left\langle A_{j}\right|  H\left|  1\right\rangle
}{\lambda_{A_{j}}}.
\end{equation}
We see that at lowest order the component of $\left|  1_{\text{full}%
}\right\rangle $ in the $\left|  A_{j}\right\rangle $ direction is independent
of the other vectors $\left|  A_{j^{\prime}}\right\rangle $. \ If $\left|
1\right\rangle $ is sufficiently close to $\left|  1_{\text{full}%
}\right\rangle $ then the limitation that only a fixed number of new basis
vectors is added in step 4 of the QSE algorithm is not relevant. \ At lowest
order in $E^{-1}$ the comparison of basis components in step 3 (in the next
iteration) is the same as if we had included all remaining vectors $\left|
A_{j}\right\rangle $ at once. \ Therefore at each update only the truly
largest components are kept and the algorithm converges to some optimal
approximation of $\left|  1_{\text{full}}\right\rangle $. \ This is consistent
with the actual performance of the algorithm as we will see in some examples
later. \ In those examples we also demonstrate that the QSE algorithm is able
to find several low energy eigenvectors simultaneously. \ The only change is
that when diagonalizing the subspace $S$ we find more than one eigenvector and
apply steps 3 and 4 of the algorithm to each of the eigenvectors.

\section{Quasi-sparsity and Anderson localization}

As the name indicates the accuracy of the quasi-sparse eigenvector method
depends on the quasi-sparsity of the low energy eigenstates in the chosen
basis. \ If the eigenvectors are quasi-sparse then the QSE method provides an
efficient way to find the important basis vectors. \ In the context of our
diagonalization/Monte Carlo approach, this means that diagonalization does
most of the work and only a small amount of correction is needed. \ This
correction is found by Monte Carlo sampling the remaining basis vectors, a
technique called stochastic error correction \cite{sec}. \ If however the
eigenvectors are not quasi-sparse then one must rely more heavily on the Monte
Carlo portion of the calculation.

The fastest and most reliable way we know to determine whether the low energy
eigenstates of a Hamiltonian are quasi-sparse with respect to a chosen basis
is to use the QSE algorithm and look at the results of the successive
iterations. \ But it is also useful to consider the question more intuitively,
and so we consider the following example.

Let $H$ be a sparse Hermitian $2000\times2000$ matrix defined by
\begin{equation}
H_{jk}=\log(j)\cdot\delta_{jk}+x_{jk}\cdot M_{jk}, \label{sym}%
\end{equation}
where $j$ and $k$ run from $1$ to $2000$, $x_{jk}$ is a Gaussian random real
variable centered at zero with standard deviation $x_{\text{rms}}=0.25$,
and\ $M_{jk}$ is a sparse symmetric matrix consisting of random $0$'s and
$1$'s such that the density of $1$'s is $5\%$. \ The reason for introducing
the $\log(j)$ term in the diagonal is to produce a large variation in the
density of states. With this choice the density of states increases
exponentially with energy. \ Our test matrix is small enough that all
eigenvectors can be found without difficulty. \ We will consider the
distribution of basis components for the eigenvectors of $H$. \ In Figure 1 we
show the square of the basis components for a given low energy eigenvector
$\left|  v\right\rangle .$ \ The basis components are sorted in order of
descending importance. \ The ratio of $\Delta E$, the average spacing between
neighboring energy levels, to $x_{\text{rms}}$ is $0.13$. \ We see that the
eigenvector is dominated by a few of its most important basis components. \ In
Figure 2 we show the same plot for another eigenstate but one where the
spacing between levels is three times smaller, $\Delta E/x_{\text{rms}%
}=0.041.$ \ This eigenvector is not nearly as quasi-sparse. \ The effect is
even stronger in Figure 3, where we show an eigenvector such that the spacing
between levels is $\Delta E/x_{\text{rms}}=0.024$.

Our observations show a strong effect of the density of states on the
quasi-sparsity of the eigenvectors. \ States with a smaller spacing between
neighboring levels tend to have basis components that extend throughout the
entire space, while states with a larger spacing tend to be quasi-sparse.
\ The relationship between extended versus localized eigenstates and the
density of states has been studied in the context of Anderson localization and
metal-insulator transitions \cite{biswas}.\ \ The simplest example is the
tight-binding model for a single electron on a one-dimensional lattice with
$Z$ sites,%

\begin{equation}
H=\sum_{j}d_{j}\left|  j\right\rangle \left\langle j\right|  +\sum
_{\left\langle jj^{\prime}\right\rangle }t_{jj^{\prime}}\left|  j\right\rangle
\left\langle j^{\prime}\right|  .
\end{equation}
$\left|  j\right\rangle $ denotes the atomic orbital state at site $j,$
$d_{j}$ is the on-site potential, and $t_{jj^{\prime}}$ is the hopping term
between nearest neighbor sites $j$ and $j^{\prime}$. \ If both terms are
uniform ($d_{j}=d,$ $t_{jj^{\prime}}=t$) then the eigenvalues and eigenvectors
of $H$ are
\begin{align}
Hv_{n}  &  =(d+2t\cos\tfrac{2\pi n}{Z})v_{n},\\
v_{n}  &  =\tfrac{1}{\sqrt{Z}}\sum_{j}e^{i\tfrac{2\pi nj}{Z}}\left|
j\right\rangle ,
\end{align}
where $n=1,\cdots,Z$ labels the eigenvectors. \ In the absence of diagonal and
off-diagonal disorder, the eigenstates of $H$ extend throughout the entire
lattice. \ The eigenvalues are also approximately degenerate, all lying within
an interval of size 4$t$. \ However, if diagonal and/or off-diagonal disorder
is introduced, the eigenvalue spectrum becomes less degenerate. \ If the
disorder is sufficiently large, the eigenstates become localized to only a few
neighboring lattice sites giving rise to a transition of the material from
metal to insulator.

We can regard a sparse quantum Hamiltonian as a similar type of system, one
with both diagonal and general off-diagonal disorder. \ If the disorder is
sufficient such that the eigenvalues become non-degenerate, then the
eigenvectors will be quasi-sparse. \ We reiterate that the most reliable way
to determine if the low energy states are quasi-sparse is to use the QSE
algorithm. \ Intuitively, though, we expect the eigenstates to be quasi-sparse
with respect to a chosen basis if the spacing between energy levels is not too
small compared with the size of the off-diagonal entries of the Hamiltonian matrix.

\section{Finite matrix examples}

As a first test of the QSE method, we will find the lowest four energy states
of the random symmetric matrix $H$ defined in (\ref{sym}). \ So that there is
no misunderstanding, we should repeat that diagonalizing a $2000\times2000$
matrix is not difficult. \ The purpose of this test is to analyze the
performance of the method in a controlled environment. \ One interesting twist
is that the algorithm uses only small pieces of the matrix and operates under
the assumption that the space may be infinite dimensional. \ A sample MATLAB
program similar to the one used here has been printed out as a tutorial
example in \cite{tutorial}.

\ The program starts from a random configuration, 70 basis states for each of
the four eigenvectors. \ With each iteration we select $10$ replacement basis
states for each of the eigenvectors. \ In Figure 4 we show the exact energies
and the results of the QSE\ method as functions of iteration number. \ In
Figure 5 we show the inner products of the normalized QSE eigenvectors with
the normalized exact eigenvectors. \ We note that all of the eigenvectors were
found after about 15 iterations and remained stable throughout successive
iterations. \ Errors are at the $5$ to $10\%$ level, which is about the
theoretical limit one can achieve using this number of basis states. \ The QSE
method has little difficulty finding several low lying eigenvectors
simultaneously because it uses the distribution of basis components for each
of the eigenvectors to determine the update process. \ This provides a
performance advantage over variational-based techniques such as stochastic
diagonalization in finding eigenstates other than the ground state. \ 

As a second test we consider a sparse non-Hermitian matrix with complex
eigenvalues. \ This type of matrix is not amenable to variational-based
methods. \ We will find the four eigenstates corresponding with eigenvalues
with the lowest real part for the random complex non-Hermitian matrix%

\begin{equation}
H_{jk}^{\prime}=(1+i\cdot c_{jk})H_{jk}.
\end{equation}
$H_{jk}$ is the same matrix used previously and$\ c_{jk}$ is a uniform random
variable distributed between $-1$ and 1. \ As before the program is started
from a random configuration, 70 basis states for each of the four
eigenvectors. \ For each iteration $10$ replacement basis vectors are selected
for each of the eigenvectors. \ In Figure 6 the exact eigenvalues and the
results of the QSE run are shown in the complex plane as functions of
iteration number. \ In Figure 7 we show the inner products of the QSE
eigenvectors with the exact eigenvectors. \ All of the eigenvectors were found
after about 20 iterations and remained stable throughout successive
iterations. \ Errors were again at about the $5$ to $10\%$ level.

\section{$\phi^{4}$ theory in $1+1$ dimensions}

We now apply the QSE method to an infinite dimensional quantum Hamiltonian.
\ We consider $\phi^{4}$ theory in $1+1$ dimensions, a system that is familiar
to us from previous studies using Monte Carlo \cite{periodic} and explicit
diagonalization \cite{spectral}. \ The Hamiltonian density for $\phi^{4}$
theory in $1+1$ dimensions has the form
\[
\mathcal{H}=\tfrac{1}{2}\left(  \tfrac{\partial\phi}{\partial t}\right)
^{2}+\tfrac{1}{2}\left(  \tfrac{\partial\phi}{\partial x}\right)  ^{2}%
+\tfrac{\mu^{2}}{2}\phi^{2}+\tfrac{\lambda}{4!}\text{:}\phi^{4}\text{:},
\]
where the normal ordering is with respect to the mass $\mu$. \ We consider the
system in a periodic box of length $2L$. \ We then expand in momentum modes
and reinterpret\ the problem as an equivalent Schr\"{o}dinger equation
\cite{periodic}. \ The resulting Hamiltonian is
\begin{align}
H  &  =-\tfrac{1}{2}%
%TCIMACRO{\dsum _{n}}%
%BeginExpansion
{\displaystyle\sum_{n}}
%EndExpansion
\tfrac{\partial}{\partial q_{-n}}\tfrac{\partial}{\partial q_{n}}+\tfrac{1}{2}%
%TCIMACRO{\dsum _{n}}%
%BeginExpansion
{\displaystyle\sum_{n}}
%EndExpansion
\left(  \omega_{n}^{2}(\mu)-\tfrac{\lambda b(\mu)}{8L}\right)  \,q_{-n}q_{n}\\
&  +\tfrac{\lambda}{4!2L}%
%TCIMACRO{\dsum _{n_{1}+n_{2}+n_{3}+n_{4}=0}}%
%BeginExpansion
{\displaystyle\sum_{n_{1}+n_{2}+n_{3}+n_{4}=0}}
%EndExpansion
q_{n_{1}}q_{n_{2}}q_{n_{3}}q_{n_{4}}\nonumber
\end{align}
where
\begin{equation}
\omega_{n}(\mu)=\sqrt{\tfrac{n^{2}\pi^{2}}{L^{2}}+\mu^{2}}%
\end{equation}
and $b(\mu)$ is the coefficient for the mass counterterm
\begin{equation}
b(\mu)=%
%TCIMACRO{\dsum _{n}}%
%BeginExpansion
{\displaystyle\sum_{n}}
%EndExpansion
\tfrac{1}{2\omega_{n}(\mu)}.
\end{equation}

It is convenient to split the Hamiltonian into free and interacting parts with
respect to an arbitrary mass $\mu^{\prime}$:%

\begin{equation}
H_{free}=-\tfrac{1}{2}%
%TCIMACRO{\dsum _{n}}%
%BeginExpansion
{\displaystyle\sum_{n}}
%EndExpansion
\tfrac{\partial}{\partial q_{-n}}\tfrac{\partial}{\partial q_{n}}+\tfrac{1}{2}%
%TCIMACRO{\dsum _{n}}%
%BeginExpansion
{\displaystyle\sum_{n}}
%EndExpansion
\omega_{n}^{2}(\mu^{\prime})\,q_{-n}q_{n},
\end{equation}%
\begin{align}
H  &  =H_{free}+\tfrac{1}{2}%
%TCIMACRO{\dsum _{n}}%
%BeginExpansion
{\displaystyle\sum_{n}}
%EndExpansion
\left(  \mu^{2}-\mu^{\prime2}-\tfrac{\lambda b(\mu)}{8L}\right)  q_{-n}q_{n}\\
&  +\tfrac{\lambda}{4!2L}%
%TCIMACRO{\dsum _{n_{1}+n_{2}+n_{3}+n_{4}=0}}%
%BeginExpansion
{\displaystyle\sum_{n_{1}+n_{2}+n_{3}+n_{4}=0}}
%EndExpansion
q_{n_{1}}q_{n_{2}}q_{n_{3}}q_{n_{4}}.\nonumber
\end{align}
$\mu^{\prime}$ is used to define the basis states of our Fock space. \ Since
$H$ is independent of $\mu^{\prime}$, we perform calculations for different
$\mu^{\prime}$ to obtain a reasonable estimate of the error. \ It is also
useful to find the range of values for $\mu^{\prime}$ which maximizes the
quasi-sparsity of the eigenvectors and therefore improves the accuracy of the
calculation. \ For the calculations presented here, we set the length of the
box to size $L=5\pi\mu^{-1}$. \ We restrict our attention to momentum modes
$q_{n}$ such that $\left|  n\right|  \leq N_{\max}$, where $N_{\max}=20$.
\ This corresponds with a momentum cutoff scale of $\Lambda=4\mu.$

To implement the QSE algorithm on this infinite dimensional Hilbert space, we
first define ladder operators with respect to $\mu^{\prime}$,
\begin{align}
a_{n}(\mu^{\prime})  &  =\tfrac{1}{\sqrt{2\omega_{n}(\mu^{\prime})}}\left[
q_{n}\omega_{n}(\mu^{\prime})+\tfrac{\partial}{\partial q_{-n}}\right] \\
a_{n}^{\dagger}(\mu^{\prime})  &  =\tfrac{1}{\sqrt{2\omega_{n}(\mu^{\prime})}%
}\left[  q_{-n}\omega_{n}(\mu^{\prime})-\tfrac{\partial}{\partial q_{n}%
}\right]  .
\end{align}
The Hamiltonian can now be rewritten as%

\begin{align}
H  &  =%
%TCIMACRO{\dsum _{n}}%
%BeginExpansion
{\displaystyle\sum_{n}}
%EndExpansion
\omega_{n}(\mu^{\prime})a_{n}^{\dagger}a_{n}+\tfrac{1}{4}(\mu^{2}-\mu
^{\prime2}-\tfrac{\lambda b}{8L})%
%TCIMACRO{\dsum _{n}}%
%BeginExpansion
{\displaystyle\sum_{n}}
%EndExpansion
\tfrac{\left(  a_{-n}+a_{n}^{\dagger}\right)  \left(  a_{n}+a_{-n}^{\dagger
}\right)  }{\omega_{n}(\mu^{\prime})}\label{ha}\\
&  +\tfrac{\lambda}{192L}%
%TCIMACRO{\dsum _{n_{1}+n_{2}+n_{3}+n_{4}=0}}%
%BeginExpansion
{\displaystyle\sum_{n_{1}+n_{2}+n_{3}+n_{4}=0}}
%EndExpansion
\left[  \tfrac{\left(  a_{n_{1}}+a_{-n_{1}}^{\dagger}\right)  }{\sqrt
{\omega_{n_{1}}(\mu^{\prime})}}\tfrac{\left(  a_{n_{2}}+a_{-n_{2}}^{\dagger
}\right)  }{\sqrt{\omega_{n_{2}}(\mu^{\prime})}}\tfrac{\left(  a_{n_{3}%
}+a_{-n_{3}}^{\dagger}\right)  }{\sqrt{\omega_{n_{3}}(\mu^{\prime})}}%
\tfrac{\left(  a_{n_{4}}+a_{-n_{4}}^{\dagger}\right)  }{\sqrt{\omega_{n_{4}%
}(\mu^{\prime})}}\right]  .\nonumber
\end{align}
In (\ref{ha}) we have omitted constants contributing only to the vacuum
energy. \ We represent any momentum-space Fock state as a string of occupation
numbers, $\left|  o_{-N_{\max}},\cdots,o_{N_{\max}}\right\rangle $, where
\begin{equation}
a_{n}^{\dagger}a_{n}\left|  o_{-N_{\max}},\cdots,o_{N_{\max}}\right\rangle
=o_{n}\left|  o_{-N_{\max}},\cdots,o_{N_{\max}}\right\rangle .
\end{equation}
From the usual ladder operator relations, it is straightforward to calculate
the matrix element of $H$ between two arbitrary Fock states.

Aside from calculating matrix elements, the only other fundamental operation
needed for the QSE algorithm is the generation of new basis vectors. \ The new
states should be connected to some old basis vector through non-vanishing
matrix elements of $H$. \ Let us refer to the old basis vector as $\left|
e\right\rangle $. \ For this example there are two types of terms in our
interaction Hamiltonian, a quartic interaction
\begin{equation}%
%TCIMACRO{\dsum _{n_{1},n_{2},n_{3}}}%
%BeginExpansion
{\displaystyle\sum_{n_{1},n_{2},n_{3}}}
%EndExpansion
\left(  a_{n_{1}}+a_{-n_{1}}^{\dagger}\right)  \left(  a_{n_{2}}+a_{-n_{2}%
}^{\dagger}\right)  \left(  a_{n_{3}}+a_{-n_{3}}^{\dagger}\right)  \left(
a_{-n_{1}-n_{2}-n_{3}}+a_{n_{1}+n_{2}+n_{3}}^{\dagger}\right)  ,
\end{equation}
and a quadratic interaction
\begin{equation}%
%TCIMACRO{\dsum _{n}}%
%BeginExpansion
{\displaystyle\sum_{n}}
%EndExpansion
\left(  a_{-n}+a_{n}^{\dagger}\right)  \left(  a_{n}+a_{-n}^{\dagger}\right)
.
\end{equation}
To produce a new vector from $\left|  e\right\rangle $ we simply choose one of
the possible operator monomials
\begin{align}
&  a_{n_{1}}a_{n_{2}}a_{n_{3}}a_{-n_{1}-n_{2}-n_{3}},\,a_{-n_{1}}^{\dagger
}a_{n_{2}}a_{n_{3}}a_{-n_{1}-n_{2}-n_{3}},\cdots,\\
&  a_{-n}a_{n},\,a_{n}^{\dagger}a_{-n}^{\dagger},\cdots\nonumber
\end{align}
and act on $\left|  e\right\rangle $. \ Our experience is that the
interactions involving the small momentum modes are generally more important
than those for the large momentum modes, a signal that the ultraviolet
divergences have been properly renormalized. \ For this reason it is best to
arrange the selection probabilities such that the smaller values of $\left|
n_{1}\right|  $, $\left|  n_{2}\right|  $, $\left|  n_{3}\right|  $ and
$\left|  n\right|  $ are chosen more often.

For each QSE iteration, $50$ new basis vectors were selected for each
eigenstate and $250$ basis vectors were retained. \ The results for the lowest
energy eigenvalues are shown in Figure 8. \ The error bars were estimated by
repeating the calculation for different values of the auxiliary mass parameter
$\mu^{\prime}$.

From prior Monte Carlo calculations we know that the theory has a phase
transition at $\frac{\lambda}{4!}\approx2.5\mu^{2}$ corresponding with
spontaneous breaking of the $\phi\rightarrow-\phi$ reflection symmetry. \ In
the broken phase there are two degenerate ground states and we refer to these
as the even and odd vacuum states. \ In Figure 8 we see signs of a second
order phase transition near $\frac{\lambda}{4!}\approx2.5\mu^{2}$. \ Since we
are working in a finite volume the spectrum is discrete, and we can track the
energy eigenvalues as functions of the coupling. \ Crossing the phase
boundary, we see that the vacuum in the symmetric phase becomes the even
vacuum in the broken phase while the one-particle state in the symmetric phase
becomes the odd vacuum. \ The energy difference between the states is also in
agreement with a Monte Carlo calculation of the same quantities. \ The state
marking the two-particle threshold in the symmetric phase becomes the
one-particle state above the odd vacuum, while the state at the three-particle
threshold becomes the one-particle state above the even vacuum. \ These
one-particle states should be degenerate in the infinite volume limit. \ One
rather unusual feature is the behavior of the first two-particle state above
threshold in the symmetric phase. \ In the symmetric phase this state lies
close to the two-particle threshold. \ But as we cross the phase boundary the
state which was the two-particle threshold is changed into a one-particle
state. \ Thus our two-particle state is pushed up even further to become a
two-particle state above the even vacuum and we see a pronounced level crossing.

We note that while the one-particle mass vanishes near the critical point, the
energies of the two-particle and three-particle thresholds reach a minimum but
do not come as close to zero energy. \ It is known that this model is
repulsive in the two-particle scattering channel. \ In a large but finite
volume the ground state and one-particle states do not feel significant finite
volume effects. \ The two-particle state at threshold, however, requires that
the two asymptotic particles be widely separated. \ In our periodic box of
length 2$L$ the maximal separation distance is $L$ and we expect an increase
in energy with respect to twice the one-particle mass of size $\sim V(L)$,
where $V$ is the potential energy between particles. \ Likewise a
three-particle state will increase in energy an amount $\sim3V(2L/3)$. \ Our
results indicate that finite volume effects for the excited states are
significant for this value of $L$.

\section{Summary}

We have proposed a new approach which combines both diagonalization and Monte
Carlo within a computational scheme. \ The motivation for our approach is to
take advantage of the strengths of the two computational methods in their
respective domains. We remedy sign and phase oscillation problems by handling
the interactions of the most important basis states exactly using
diagonalization, and we deal with storage and CPU problems by stochastically
sampling the contribution of the remaining states. \ We discussed the
diagonalization part of the method in this paper. \ The goal of
diagonalization within our scheme is to find the most important basis vectors
of the low energy eigenstates and treat the interactions among them exactly.
\ We have introduced a new diagonalization method called quasi-sparse
eigenvector diagonalization which achieves this goal efficiently and can
operate using any basis, either orthogonal or non-orthogonal, and any sparse
Hamiltonian, either real, complex, Hermitian, non-Hermitian,
finite-dimensional, or infinite-dimensional. \ Quasi-sparse eigenvector
diagonalization is the only method we know which can address all of these problems.

We considered three examples which tested the performance of the algorithm.
\ We found the lowest energy eigenstates for a random sparse real symmetric
matrix, the lowest eigenstates (sorted according to the real part of the
eigenvalue) for a random sparse complex non-Hermitian matrix, and the lowest
energy eigenstates for an infinite-dimensional Hamiltonian defined by $1+1$
dimensional $\phi^{4}$ theory in a periodic box.

We regard QSE diagonalization as only a starting point for the Monte Carlo
part of the calculation. \ Once the most important basis vectors are found and
their interactions treated exactly, a technique called stochastic error
correction is used to sample the contribution of the remaining basis vectors.
\ This method is introduced in \cite{sec}.

\paragraph*{Acknowledgments}

We thank P. van Baal, H. S. Seung, H. Sompolinsky, and M. Windoloski for
useful discussions. \ Support provided by the National Science Foundation.

\begin{figure}[t]
\begin{center}
\epsfxsize=20pc \epsfbox{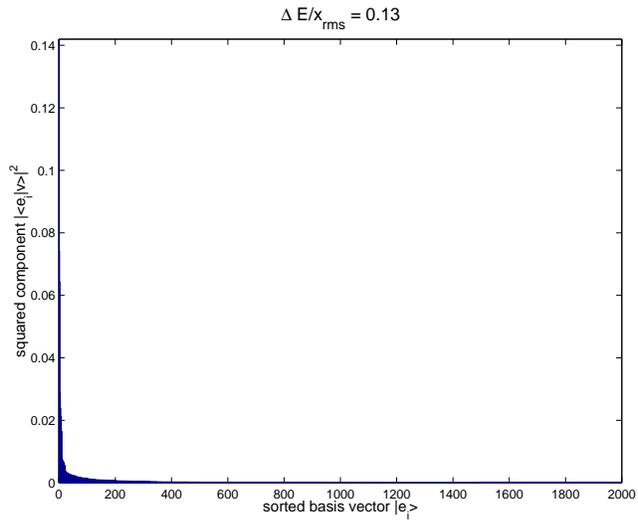}
%postscript image file name
%
%
%
%
%
%
%
\end{center}
\caption{Distribution of basis components for an eigenvector where the spacing
between consecutive levels is $\Delta E=0.13x_{\text{rms}}$. }%
\end{figure}

\begin{figure}[t]
\begin{center}
\epsfxsize=20pc \epsfbox{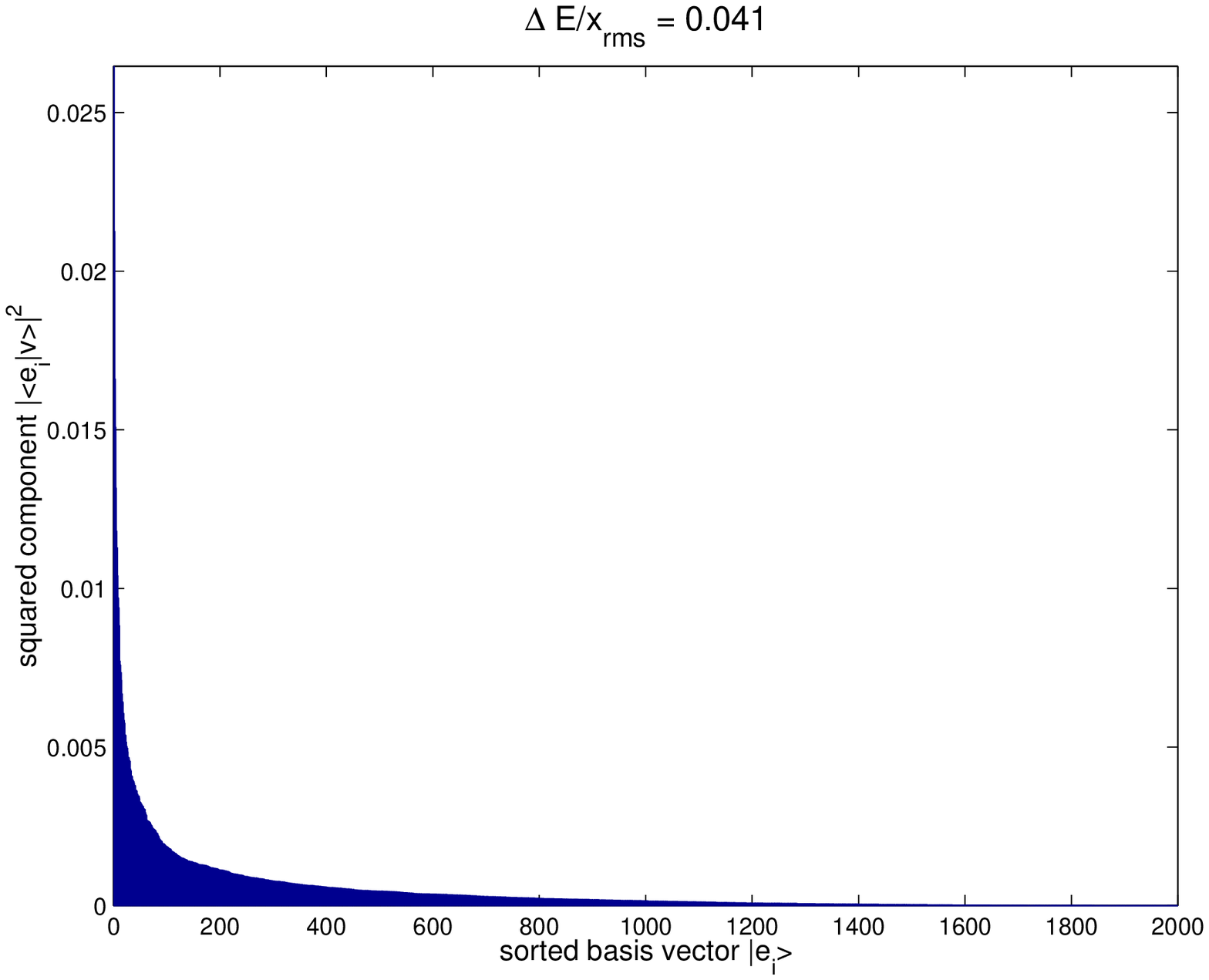}
\end{center}
\caption{Distribution of basis components for an eigenvector where $\Delta
E=0.041x_{\text{rms}}$. }%
\end{figure}

\begin{figure}[t]
\begin{center}
\epsfxsize=20pc \epsfbox{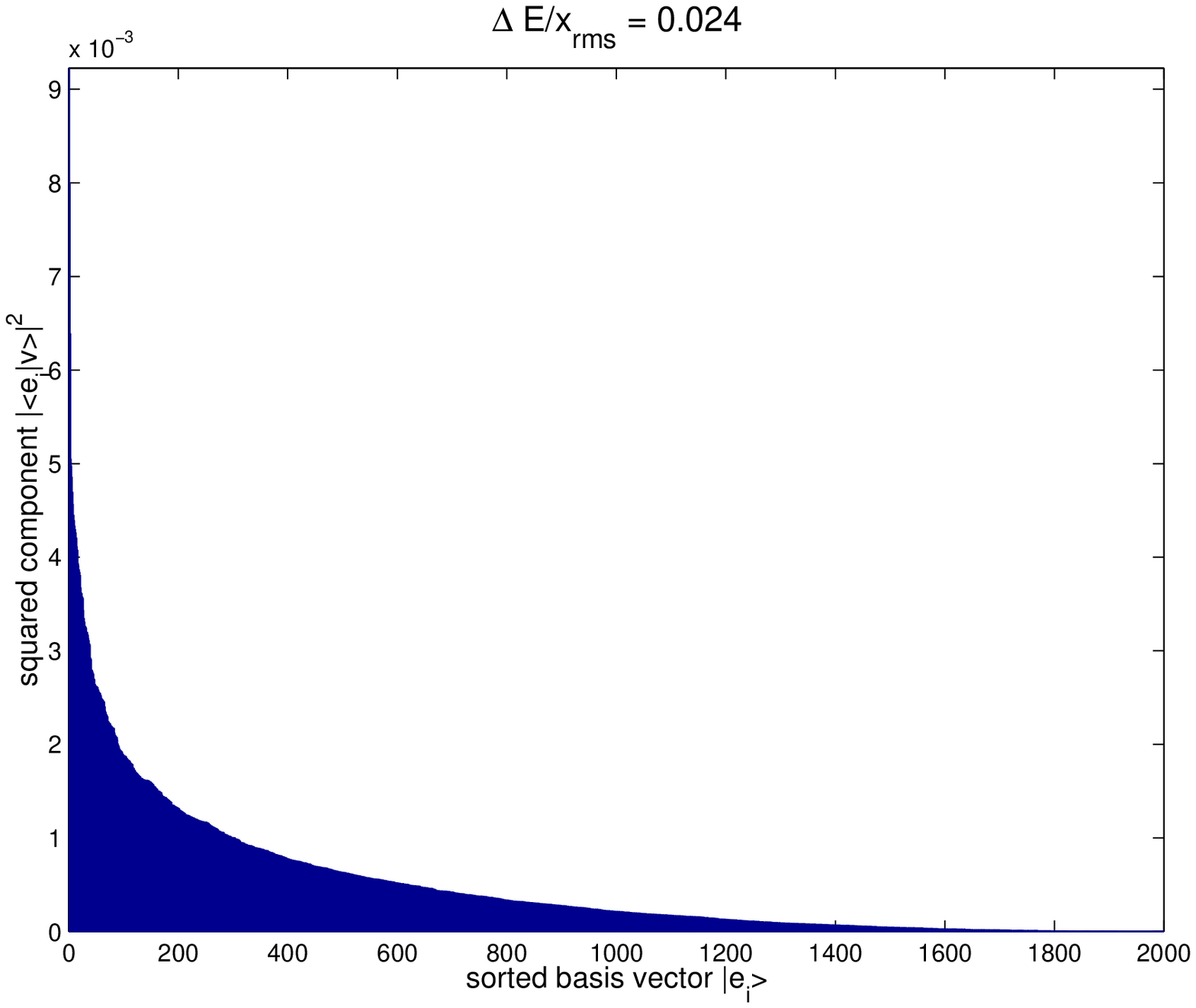}
\end{center}
\caption{Distribution of basis components for an eigenvector where $\Delta
E=0.024x_{\text{rms}}$. }%
\end{figure}

\begin{figure}[t]
\begin{center}
\epsfxsize=20pc \epsfbox{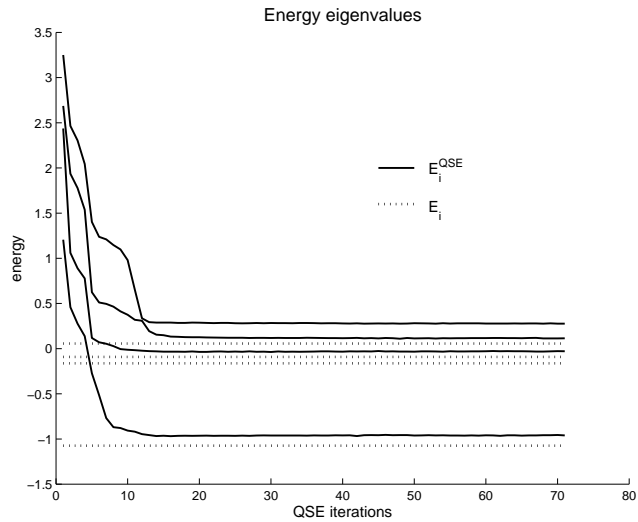}
\end{center}
\caption{Comparison of the four lowest exact energies $E_{i}$ and QSE results
$E_{i}^{\text{QSE}}$ as functions of iteration number. }%
\end{figure}

\begin{figure}[t]
\begin{center}
\epsfxsize=20pc \epsfbox{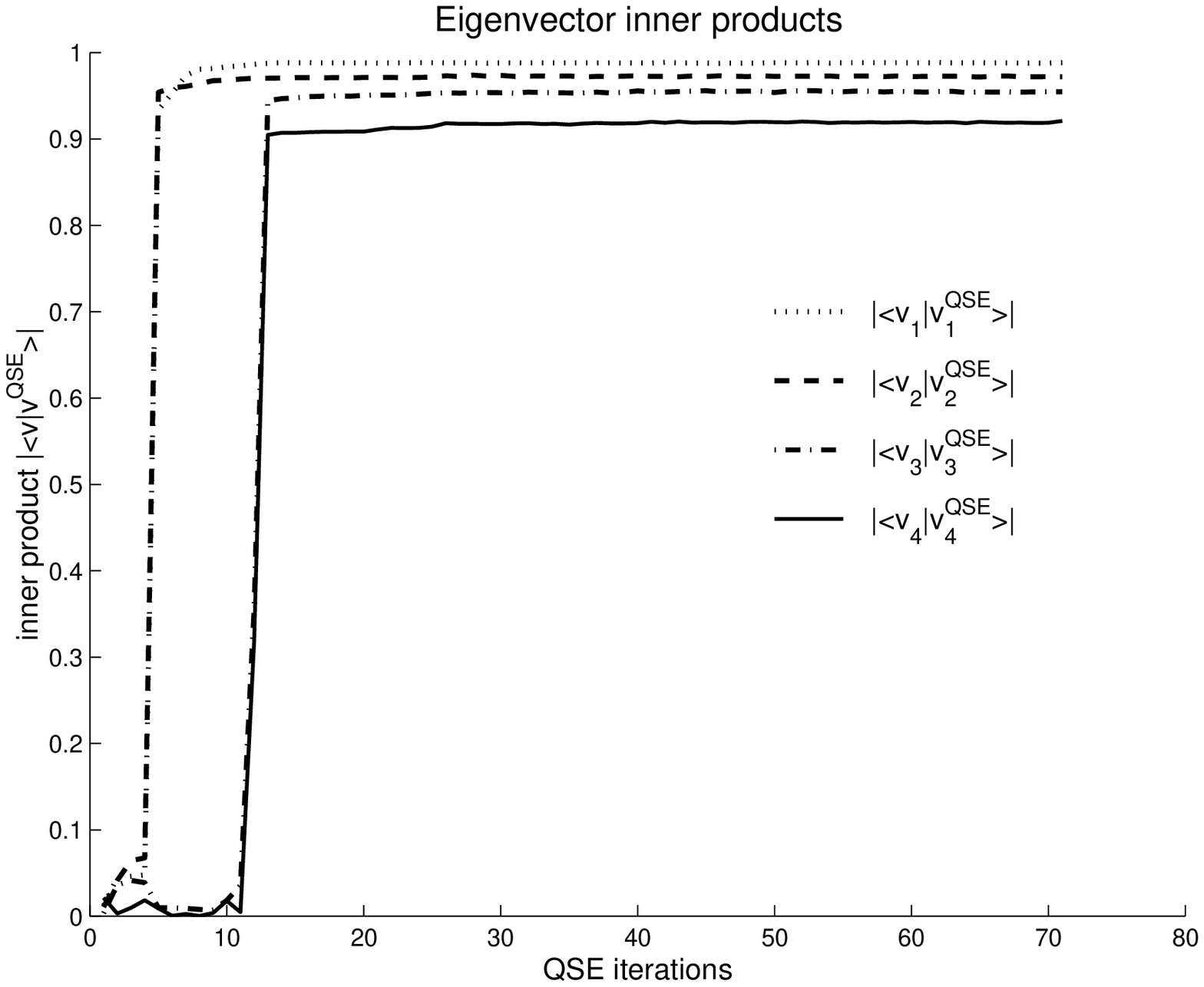}
\end{center}
\caption{Inner products between the normalized exact eigenvectors $\left|
v_{i}\right\rangle $ and the QSE results $\left|  v_{i}^{\text{QSE}%
}\right\rangle $ as functions of iteration number. }%
\end{figure}

\begin{figure}[t]
\begin{center}
\epsfxsize=20pc \epsfbox{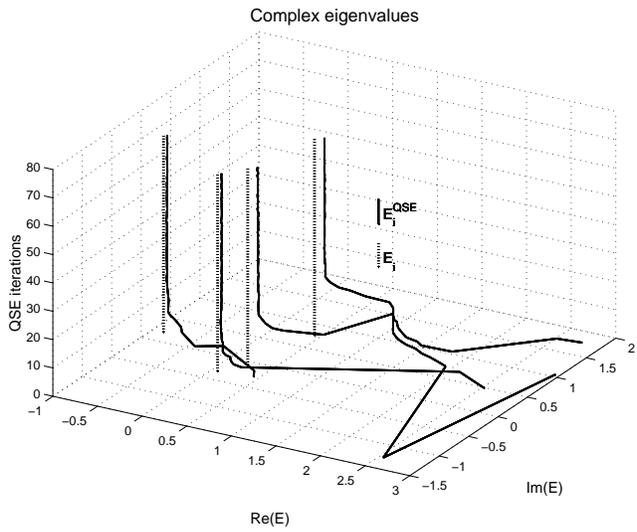}
\end{center}
\caption{Comparison of the four lowest exact eigenvalues $E_{i}$ (sorted by
real part) and QSE results $E_{i}^{\text{QSE}}$ in the complex plane as
functions of iteration number. }%
\end{figure}

\begin{figure}[t]
\begin{center}
\epsfxsize=20pc \epsfbox{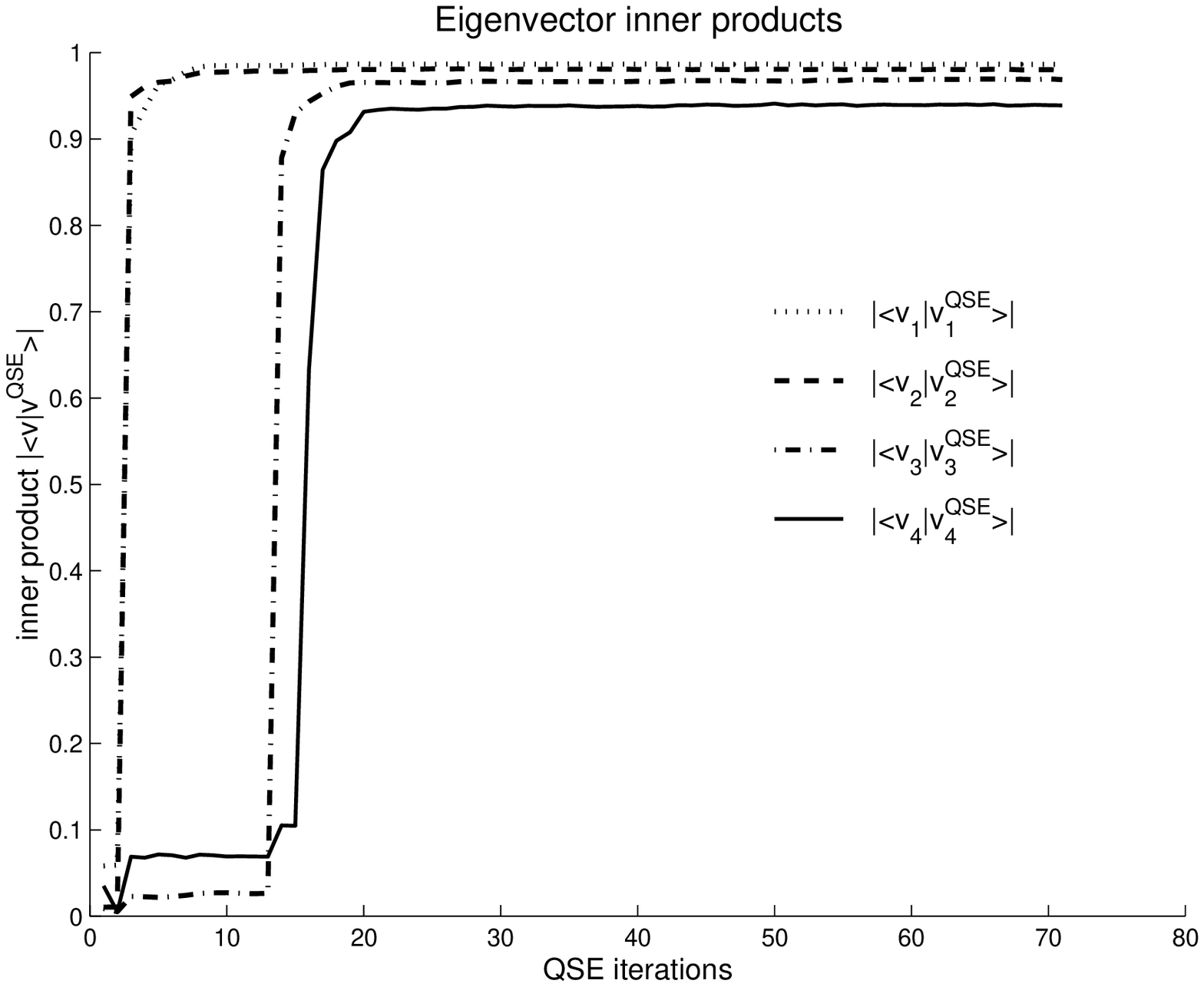}
\end{center}
\caption{Inner products between the normalized exact eigenvectors $\left|
v_{i}\right\rangle $ and the QSE results $\left|  v_{i}^{\text{QSE}%
}\right\rangle $ as functions of iteration number. }%
\end{figure}

\begin{figure}[t]
\begin{center}
\epsfxsize=20pc \epsfbox{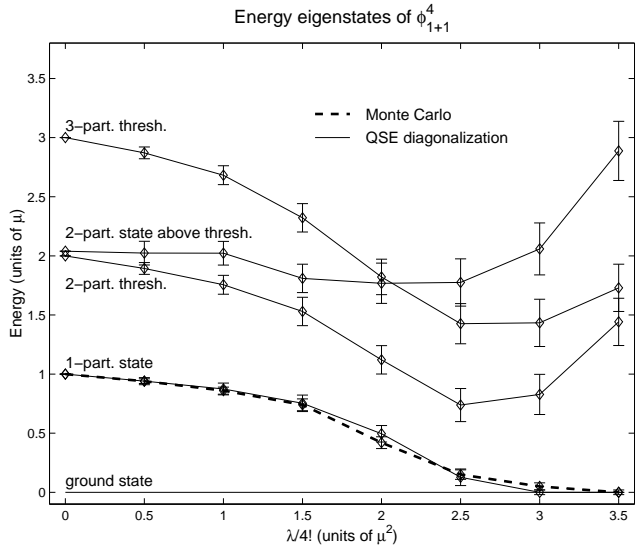}
\end{center}
\caption{Energy eigenvalues of $\phi_{1+1}^{4}$ as functions of the coupling
constant. }%
\end{figure}
\end{document}